\documentclass{appolb}
\usepackage{graphicx}

\begin{document}
\title{Parton Distribution Functions and Their Impact on Precision of the Current Theory Calculations
\thanks{Presented at XXX Cracow Epiphany Conference on Precision Physics at High Energy Colliders - dedicated to the memory of Staszek Jadach}%
}
\author{Maria Ubiali
\address{DAMTP, University of Cambridge, Wilberforce Road, Cambridge, CB3 0WA, UK}}

\maketitle
\begin{abstract}
The unprecedented precision of experimental measurements at the Large Hadron Collider (LHC) and the increased statistics that will be reached in the High-Luminosity phase of the LHC (HL-LHC) are pushing the phenomenology community to a new precision frontier, in which new challenges present themselves and new questions arise. A key ingredients of theoretical predictions at hadron colliders are the Parton Distribution Functions (PDFs) of the proton. This contribution highlights some of the new developments in the determination of PDFs from a global set of experimental data, from approximate N$^3$LO PDFs and the inclusion of theory uncertainties in PDF fits, to the realisation of the non trivial interplay between parton densities at large-$x$ and possible signals of New Physics in high energy tails of the distributions, which highlights the synergy between high energy and low energy experimental programs.\end{abstract}
  
\section{Introduction}
Precision phenomenology at hadron colliders and searches for deviations from the Standard Model (SM) predictions rely upon a precise and accurate estimate of the uncertainty in the SM predictions. Most theory predictions for hadron colliders require Parton Distribution Functions (PDFs) \cite{PDF4LHCWorkingGroup:2022cjn,Ball:2017nwa,Hou:2019efy,Bailey:2020ooq,Ball:2021leu,ATLAS:2021vod,Accardi:2016qay,Alekhin:2017kpj,Butterworth:2015oua}, the non-perturbative functions parametrising the subnuclear structure of the protons in terms of their partonic constituents for high-energy scattering processes. 

PDFs cannot be obtained by perturbative methods, hence they must be fitted from experimental data. The space of possible parton distributions is an infinite-dimensional space of functions obeying the DGLAP evolution equations; as such, given only finite amounts of data, it is an ill-posed problem to determine the PDFs. Therefore, any PDF fitting attempt must initially restrict the infinite-dimensional PDF space to a finite-dimensional space instead, by assuming a functional form for the PDFs at some initial scale $Q_0$\footnote{This can be a set of polynomials, that typically involves 30-50 parameters or a redundant parametrization such as a deep neural network, involving hundreds of parameters and a regularisation or minimisation stopping criterion, depending on the methodology.}, and then DGLAP evolution is used to obtain the PDF at all scales. 

A faithful determination of PDFs and of their uncertainty is a highly non trivial problem and the precision of experimental data is presenting new challenges to PDF fitting collaborations. Accurate and precise PDFs are crucial at the LHC. For example, despite the progress in the precise determination of PDFs, PDF uncertainty is still one of the largest sources of theoretical uncertainty affecting the predictions for Higgs boson production~\cite{LHCHiggsCrossSectionWorkingGroup:2016ypw,Amoroso:2022eow,Baglio:2022wzu}.  The experimental PDF uncertainty is not the only component of the total uncertainty, as there is another component representing the mismatch in the perturbative order of the PDFs, evaluated at N$^2$LO, and the perturbative QCD cross sections evaluated at N$^3$LO~\cite{Anastasiou:2016cez,Mistlberger:2018etf}. The recent publication of approximate N$^3$LO PDF sets~\cite{Cridge:2023ryv,Cridge:2024exf,NNPDF:2024nan}, which will be discussed in Sect.~\ref{sec:challenges}, is a crucial step in reducing this source of uncertainty. 
PDFs do also contribute to precise extraction of the SM parameters from the LHC data, such as for example the strong coupling constant~\cite{ATLAS:2023lhg} and the $W$ boson mass~\cite{ATLAS:2024erm}, for which different PDF sets yield results that are significantly different as compared to the size of statistical and systematic uncertainties. 

Finally, PDF uncertainties become sizeable at large-$x$ due to the lack of precise experimental constraints in that region, and this yields a large uncertainty in the predictions for the high-energy tails of the measured distributions, where programs of indirect new physics searches focus, see for example the discussion in Ref.~\cite{Ball:2022qtp}. A precise and accurate description of the large-$x$ content of the proton has always been crucial. As an illustration, the discrepancy that was observed by the CDF collaboration in 1995 between the SM predictions for jet inclusive cross section at large $p_T$~\cite{CDF:1995odg} was mostly due to the lack of an estimate of the uncertainty of the gluon at large-$x$, and indeed the discrepancy disappeared once the CDF jet distributions were included by CTEQ in their global analysis of PDFs~\cite{Huston:1995tw}. The story taught us the importance of propagating the experimental uncertainty in the space of PDFs. However, now that uncertainties are much more carefully estimated, how can we make sure that potential deviations from the SM predictions in the high energy tails of the distributions are genuinely due to new physics rather than to our poor understanding of the proton structure at large $x$? This question will be on of the focus of the discussion in  Sect.~\ref{sec:challenges}.

\section{New frontiers and challenges}
\label{sec:challenges}
As data precision increases, so does the challenge of balancing precision and accuracy. New PDF analyses sometimes lead to significant shifts from previous sets, potentially larger than nominal uncertainties, yet this doesn't undermine accuracy if the source of shifts is identified and the fitting process is controlled.
Growing luminosity in experimental data poses challenges, especially with increasingly correlated systematics dominating datasets. Stabilizing covariance matrices and addressing systematic decorrelation are ongoing tasks crucial for assessing the goodness of fit~\cite{Bailey:2019yze,Kassabov:2022pps}.
Continuous improvements and rigorous tests, ensure robustness, particularly important for modern PDF sets, see for example discussion on closure tests in Ref.~\cite{DelDebbio:2021whr} and future tests in Ref.~\cite{Cruz-Martinez:2021rgy}.

In what follows we will select three key aspects: the theory framework advancements that is behind approximate N$^3$LO PDFs (discussed in Sect.~\ref{subsec:n3lo}), the inclusion of theoretical uncertainty in PDF fits (discussed in Sect.~\ref{subsec:mhou}), and finally the interplay between PDF and new physics (discussed in Sect.~\ref{subsec:np}).

\subsection{Approximate N$^3$LO PDFs}
\label{subsec:n3lo}
Calculations of partonic cross sections at N$^3$LO order have been available for a long time for massless deep-inelastic scattering (DIS), and have become recently available for an increasingly large set of hadron processes, see Ref.~\cite{Caola:2022ayt} for an overview. To obtain theoretical predictions for hadronic observables at the same level of accuracy, the N$^3$LO partonic cross sections must be convoluted with PDFs determined at the same perturbative order. The main bottleneck in carrying out this programme is the lack of exact expressions for the N$^3$LO splitting functions that govern the scale dependence of the PDFs: for these only partial information is available, see Ref.~\cite{NNPDF:2024nan} for a review. However, a combination of the available partial results is possible, and an approximate determination of the N$^3$LO splitting functions, see for example Refs.~\cite{Moch:2023tdj,Cridge:2021pxm}.

Currently, two approximate N$^3$LO PDF sets are available, with the MSHT set~\cite{Cridge:2021pxm} available since a couple of year and the combined QED and N$^3$LO MSHT set~\cite{Cridge:2023ryv} and 
the NNPDF4.0 set~\cite{NNPDF:2024nan} become available only recently. 
There are several differences in the approaches, as the {\tt MSHTaN3LO} set includes all available theory input at the time of publication, while the {\tt NNPDF40aN3LO} includes more theory inputs that were published in between, in particular 6(1) extra momentum for the $P_{qg}$, $P_{qq}$ ($P_{gq}, \,P_{gg}$) splitting functions, some terms in the large $n_f$ limit, and several sub-leading small-$x$ and large-$x$ terms. A benchmark between the two approaches at the level of splitting functions is in progress in a Les Houches benchmark exercise. Another important difference is that in the MSHT approach the incomplete N$^3$LO terms are added as variation in the prior and estimated by fitting nuisance parameters to the data, hence the posterior is determined by fitting the data, while the NNPDF approximation incorporates only theory inputs and their variations is added via an additional theory covariance matrix associated with incomplete missing higher orders (IHOU). Moreover in the NNPDF approach the missing higher order uncertainties (MHOU) associated with the NNLO contributions are added via a MHOU theory covariance matrix~\cite{AbdulKhalek:2019mps,AbdulKhalek:2019mzd}. 

While the differences in the approach yield slightly discrepant predictions, overall a good perturbative convergence is observed, with differences decreasing as the perturbative order increases, and the approximate N$^3$LO result always compatible with the NNLO within uncertainties.
Also, while for quark PDFs the difference between NNLO and aN$^3$LO results is extremely small, for the gluon PDF a more significant shift is observed between NNLO and N$^3$LO (this is much more marked in the MSHT set than in the NNPDF set), thus making the inclusion of N$^3$LO crucial for precision phenomenology.

\subsection{Missing higher order uncertainties}
\label{subsec:mhou}

As the overall precision in the determination grows, it is paramount to consistently treat the theoretical uncertainties of the input predictions in PDF fits. Until very recently PDF uncertainties in NNLO PDF fits only included experimental component. However such fits are performed by comparing fixed-order theoretical predictions to experimental data and theoretical predictions are affected by MHOU, which is usually estimated by means of scale variation. 
The inclusion of scale variations in a state-of-the-art NLO PDF fit was presented for the first time in Refs.~\cite{AbdulKhalek:2019mps,AbdulKhalek:2019mzd}. The approach,  built upon the construction of a theory covariance matrix that describes the scale variations of the processes included in a PDF fit and models their theoretical correlations, has now been generalised to a NNLO PDF set~\cite{NNPDF:2024dpb}. The theory covariance matrix so determined can be added to the experimental covariance matrix, and the data are fitted using a total covariance matrix, which is the sum of the two contributions. The inclusion of MHOUs generally improves perturbative convergence and represents an important step towards achieving accuracy and improving perturbative convergence. 
A similar method has been also been applied to nuclear uncertainties, which are relevant in the description of data based on nuclear targets~\cite{Ball:2020xqw,Ball:2018twp}.

With the availability of PDF sets including MHOUs the total uncertainty of any theory prediction would combine PDF uncertainty (now including a MHOU component), with the MHOU uncertainty on the partonic cross section computations. In order to keep into account the correlation between the scale variations in the process used for PDF determination, and the scale variation on the partonic cross section that an external user wishes to compute, one should compute the cross-correlation of MHO uncertainties between the predicted process and those used for PDF determination~\cite{Harland-Lang:2018bxd,Ball:2021icz}.
An alternative option for the inclusion of MHOUs on predictions that accounts for MHOUs on PDFs, and their full correlation to MHOUs on the partonic cross sections, is to include the scale variation in the Monte Carlo sampling, as proposed in the {\tt MCscales} approach~\cite{Kassabov:2022orn}. Basically the Monte Carlo sampling method used to propagate the experimental uncertainty of the data into the PDFs is extended by 
adding scale fluctuations for each theoretical prediction on top of the fluctuations on the
input data that are used to propagate experimental uncertainties.
The development and the comparison of the various ways of including MHOUs in the estimate of PDF uncertainties will become increasingly crucial as the experimental component of PDF uncertainties is set to decrease thanks to the increased statistics and better understanding of the experimental systematics in Run III and IV, as well as in the HL-LHC phase.

\subsection{PDFs and new physics interplay}
\label{subsec:np}
Recent PDF analyses indicate that the LHC data are increasingly crucial in pinning down the parton densities, especially for the gluon and sea quarks the intermediate to large-$x$ regions and its constraining power will become even more crucial in the HL-LHC phase~\cite{AbdulKhalek:2018rok}. 
As an increasing number of high-energy data from the LHC are included in PDF fits, the tails of the distributions that are used in PDF determination are potentially affected by new physics effects, which are usually parametrised by means of a model independent EFT expansion~\cite{Brivio:2017vri}. To make sure that new physics is not absorbed or “fitted away” in the PDFs, one would either have to exclude these data, thus losing potentially important constraints, or carefully disentangle the SM and new physics effects.

Several studies have been performed in this direction by both experimental collaborations and theory groups~\cite{ZEUS:2019cou,Carrazza:2019sec,CMS:2021yzl,Greljo:2021kvv,McCullough:2022hzr,Gao:2022srd,Kassabov:2023hbm,Hammou:2023heg,Costantini:2024xae,Shen:2024uop}. In particular, it has been shown that the determination of SMEFT Wilson coefficients from a fit of LHC data, like the determination of SM precision parameters from LHC data~\cite{Forte:2020pyp}, might display a non-negligible interplay with the input set of PDFs used to compute theory predictions. For example, in~\cite{Carrazza:2019sec} it was shown that, while the effect of four-fermion operators on Deep Inelastic Scattering (DIS) data can be non-negligible, if DIS data were fitted while taking the effect of such operators into account, the fit quality would deteriorate proportionally to the energy scale of the data included in the determination. This makes a fit based on DIS-only data "safe" from new physics effects. 

On the other hand in the context of high-mass Drell-Yan, especially in the HL-LHC scenario, neglecting the cross-talk between the large-$x$ PDFs and the SMEFT effects in the tails could potentially miss new physics manifestations or misinterpret them~\cite{Greljo:2021kvv,Hammou:2023heg,Iranipour:2022iak}, as the bounds on SMEFT operators are significantly broader if the PDFs are fitted by including the effect of the $\hat{W}$ and $\hat{Y}$ universal operators~\cite{Torre:2020aiz} in the high energy tails of the data. In Ref.~\cite{Hammou:2023heg} it was shown that the large-$x$ antiquark distributions can absorb the effect of universal new physics in the tails of the Drell-Yan distributions by leading to significantly softer antiquark distributions in the large-$x$ region, far beyond the nominal PDF uncertainties. The study shows that the only way to avoid this kind of BSM contamination would be to  include more low-energy observables in a PDF fit, thus highlighting the synergy between low- and high-energy experiments. 

Furthermore, the analysis of the top quark sector of Ref.~\cite{Kassabov:2023hbm} demonstrates that in that case the bounds of the operators are not broadened by the interplay with the PDFs, however the correlation between the top sector and the gluon is manifest in the gluon PDF itself, which becomes softer in the large-$x$ region if the PDF fit is augmented by the top data and PDFs are fitted simultaneously along the Wilson coefficients that determine the top quark sector. This hints to a non-negligible interplay between large-$x$ gluons and hints for new physics signals in the large invariant mass tails of the $t\bar{t}$ distributions, and possibly in the high-$E_T$ jets distributions. 

The recently publicly released tool {\tt SIMUnet}~\cite{Costantini:2024xae} allows users to assess the interplay between PDFs and new physics, either by performing simultaneous fits of SMEFT operators (or operators in any other EFT expansion) alongside the PDFs of the proton, or by injecting any new physics model in the data and checking whether a global fit of PDFs can absorb the effects induced by such a model in the data.
{\tt SIMUnet}, based on the first tagged version of the public NNPDF code~\cite{NNPDF:2021uiq}, augments it with a number of new features that allow the exploration of the correlation between a fit of PDFs and BSM degrees of freedom. The tool allows a global analysis of the SMEFT combining the Higgs, top, diboson and electroweak sectors, and more data can be added in a rather straightforward way to combine the sectors presented here alongside Drell-Yan and flavour observables, for example. Further exploration of the SMEFT and PDF interplay and of the possible signal of new physics that might be fitted away by PDFs especially in the top and jets sector would add important pieces of the mosaic that has just started to be unveiled. 

\section{Conclusions}
The success of the ambitious programme of the upcoming Run III at the LHC and its subsequent High- Luminosity (HL-LHC) upgrade relies not only on achieving the highest possible accuracy in the experimental measurements and in the corresponding theoretical predictions, but also on the availability of statistically robust tools capable of yielding global interpretations of all subtle deviations from the Standard Model (SM) that the data might indicate. 

In this contribution we discussed how the careful inspection and use of novel theoretical advances such as the recently developed approximate N$^3$LO PDFs, the inclusion of MHOU and of QED effects are important to achieve the best accuracy at the LHC. Thanks to the number of newly released data, a powerful testing ground of the generalisation and extrapolation of the various PDF sets is available to be explored. Moreover the new light shed on the interplay between PDFs and new physics can further advance the synergy between new experiments of the deeply-inelastic scattering (DIS), for example at the Electron-Ion Collider~\cite{AbdulKhalek:2021gbh}, or the perspective Large Hadron-Electron Collide~\cite{LHeC:2020van} or even at the future Forward Physics Facilities planned at CERN~\cite{Cruz-Martinez:2023sdv}. 
The challenges that we are presented with are certainly a great opportunity to refine our tools and advance in our understanding of fundamental physics.
\bibliographystyle{utphys}

\begin{thebibliography}{10}

\bibitem{PDF4LHCWorkingGroup:2022cjn}
{\bfseries PDF4LHC Working Group} Collaboration, R.~D. Ball {\em et~al.},
  ``{The PDF4LHC21 combination of global PDF fits for the LHC Run III},''
  \href{http://dx.doi.org/10.1088/1361-6471/ac7216}{{\em J. Phys. G} {\bfseries
  49} no.~8, (2022) 080501}, \href{http://arxiv.org/abs/2203.05506}{{\ttfamily
  arXiv:2203.05506 [hep-ph]}}.

\bibitem{Ball:2017nwa}
{\bfseries NNPDF} Collaboration, R.~D. Ball {\em et~al.}, ``{Parton
  distributions from high-precision collider data},''
  \href{http://dx.doi.org/10.1140/epjc/s10052-017-5199-5}{{\em Eur. Phys. J.}
  {\bfseries C77} no.~10, (2017) 663},
\href{http://arxiv.org/abs/1706.00428}{{\ttfamily arXiv:1706.00428 [hep-ph]}}.

\bibitem{Hou:2019efy}
T.-J. Hou {\em et~al.}, ``{New CTEQ global analysis of quantum chromodynamics
  with high-precision data from the LHC},''
  \href{http://dx.doi.org/10.1103/PhysRevD.103.014013}{{\em Phys. Rev. D}
  {\bfseries 103} no.~1, (2021) 014013},
  \href{http://arxiv.org/abs/1912.10053}{{\ttfamily arXiv:1912.10053
  [hep-ph]}}.

\bibitem{Bailey:2020ooq}
S.~Bailey, T.~Cridge, L.~A. Harland-Lang, A.~D. Martin, and R.~S. Thorne,
  ``{Parton distributions from LHC, HERA, Tevatron and fixed target data:
  MSHT20 PDFs},'' \href{http://dx.doi.org/10.1140/epjc/s10052-021-09057-0}{{\em
  Eur. Phys. J. C} {\bfseries 81} no.~4, (2021) 341},
  \href{http://arxiv.org/abs/2012.04684}{{\ttfamily arXiv:2012.04684
  [hep-ph]}}.

\bibitem{Ball:2021leu}
{\bfseries NNPDF} Collaboration, R.~D. Ball {\em et~al.}, ``{The path to proton
  structure at 1\% accuracy},''
  \href{http://dx.doi.org/10.1140/epjc/s10052-022-10328-7}{{\em Eur. Phys. J.
  C} {\bfseries 82} no.~5, (2022) 428},
  \href{http://arxiv.org/abs/2109.02653}{{\ttfamily arXiv:2109.02653
  [hep-ph]}}.

\bibitem{ATLAS:2021vod}
{\bfseries ATLAS} Collaboration, G.~Aad {\em et~al.}, ``{Determination of the
  parton distribution functions of the proton using diverse ATLAS data from
  $pp$ collisions at $\sqrt{s} = 7$, 8 and 13~TeV},''
  \href{http://dx.doi.org/10.1140/epjc/s10052-022-10217-z}{{\em Eur. Phys. J.
  C} {\bfseries 82} no.~5, (2022) 438},
  \href{http://arxiv.org/abs/2112.11266}{{\ttfamily arXiv:2112.11266
  [hep-ex]}}.

\bibitem{Accardi:2016qay}
A.~Accardi, L.~T. Brady, W.~Melnitchouk, J.~F. Owens, and N.~Sato,
  ``{Constraints on large-$x$ parton distributions from new weak boson
  production and deep-inelastic scattering data},''
  \href{http://dx.doi.org/10.1103/PhysRevD.93.114017}{{\em Phys. Rev. D}
  {\bfseries 93} no.~11, (2016) 114017},
  \href{http://arxiv.org/abs/1602.03154}{{\ttfamily arXiv:1602.03154
  [hep-ph]}}.

\bibitem{Alekhin:2017kpj}
S.~Alekhin, J.~Blümlein, S.~Moch, and R.~Placakyte, ``{Parton distribution
  functions, $\alpha_s$, and heavy-quark masses for LHC Run II},''
  \href{http://dx.doi.org/10.1103/PhysRevD.96.014011}{{\em Phys. Rev.}
  {\bfseries D96} no.~1, (2017) 014011},
\href{http://arxiv.org/abs/1701.05838}{{\ttfamily arXiv:1701.05838 [hep-ph]}}.

\bibitem{Butterworth:2015oua}
J.~Butterworth {\em et~al.}, ``{PDF4LHC recommendations for LHC Run II},''
  \href{http://dx.doi.org/10.1088/0954-3899/43/2/023001}{{\em J. Phys. G}
  {\bfseries 43} (2016) 023001},
  \href{http://arxiv.org/abs/1510.03865}{{\ttfamily arXiv:1510.03865
  [hep-ph]}}.

\bibitem{LHCHiggsCrossSectionWorkingGroup:2016ypw}
{\bfseries LHC Higgs Cross Section Working Group} Collaboration, D.~de~Florian
  {\em et~al.}, ``{Handbook of LHC Higgs Cross Sections: 4. Deciphering the
  Nature of the Higgs Sector},''
  \href{http://arxiv.org/abs/1610.07922}{{\ttfamily arXiv:1610.07922
  [hep-ph]}}.

\bibitem{Amoroso:2022eow}
S.~Amoroso {\em et~al.}, ``{Snowmass 2021 Whitepaper: Proton Structure at the
  Precision Frontier},''
  \href{http://dx.doi.org/10.5506/APhysPolB.53.12-A1}{{\em Acta Phys. Polon. B}
  {\bfseries 53} no.~12, (2022) 12--A1},
  \href{http://arxiv.org/abs/2203.13923}{{\ttfamily arXiv:2203.13923
  [hep-ph]}}.

\bibitem{Baglio:2022wzu}
J.~Baglio, C.~Duhr, B.~Mistlberger, and R.~Szafron, ``{Inclusive production
  cross sections at N$^{3}$LO},''
  \href{http://dx.doi.org/10.1007/JHEP12(2022)066}{{\em JHEP} {\bfseries 12}
  (2022) 066}, \href{http://arxiv.org/abs/2209.06138}{{\ttfamily
  arXiv:2209.06138 [hep-ph]}}.

\bibitem{Anastasiou:2016cez}
C.~Anastasiou, C.~Duhr, F.~Dulat, E.~Furlan, T.~Gehrmann, F.~Herzog,
  A.~Lazopoulos, and B.~Mistlberger, ``{High precision determination of the
  gluon fusion Higgs boson cross-section at the LHC},''
  \href{http://dx.doi.org/10.1007/JHEP05(2016)058}{{\em JHEP} {\bfseries 05}
  (2016) 058}, \href{http://arxiv.org/abs/1602.00695}{{\ttfamily
  arXiv:1602.00695 [hep-ph]}}.

\bibitem{Mistlberger:2018etf}
B.~Mistlberger, ``{Higgs boson production at hadron colliders at N$^{3}$LO in
  QCD},'' \href{http://dx.doi.org/10.1007/JHEP05(2018)028}{{\em JHEP}
  {\bfseries 05} (2018) 028}, \href{http://arxiv.org/abs/1802.00833}{{\ttfamily
  arXiv:1802.00833 [hep-ph]}}.

\bibitem{Cridge:2023ryv}
T.~Cridge, L.~A. Harland-Lang, and R.~S. Thorne, ``{Combining QED and
  Approximate N${}^3$LO QCD Corrections in a Global PDF Fit: MSHT20qed\_an3lo
  PDFs},'' \href{http://arxiv.org/abs/2312.07665}{{\ttfamily arXiv:2312.07665
  [hep-ph]}}.

\bibitem{Cridge:2024exf}
T.~Cridge, L.~A. Harland-Lang, and R.~S. Thorne, ``{A first determination of
  the strong coupling $\alpha_S$ at approximate N$^{3}$LO order in a global PDF
  fit},'' \href{http://arxiv.org/abs/2404.02964}{{\ttfamily arXiv:2404.02964
  [hep-ph]}}.

\bibitem{NNPDF:2024nan}
{\bfseries NNPDF} Collaboration, R.~D. Ball {\em et~al.}, ``{The Path to
  N$^3$LO Parton Distributions},''
  \href{http://arxiv.org/abs/2402.18635}{{\ttfamily arXiv:2402.18635
  [hep-ph]}}.

\bibitem{ATLAS:2023lhg}
{\bfseries ATLAS} Collaboration, G.~Aad {\em et~al.}, ``{A precise
  determination of the strong-coupling constant from the recoil of $Z$ bosons
  with the ATLAS experiment at $\sqrt{s} = 8$ TeV},''
  \href{http://arxiv.org/abs/2309.12986}{{\ttfamily arXiv:2309.12986
  [hep-ex]}}.

\bibitem{ATLAS:2024erm}
{\bfseries ATLAS} Collaboration, G.~Aad {\em et~al.}, ``{Measurement of the
  W-boson mass and width with the ATLAS detector using proton-proton collisions
  at $\sqrt{s}$ = 7 TeV},'' \href{http://arxiv.org/abs/2403.15085}{{\ttfamily
  arXiv:2403.15085 [hep-ex]}}.

\bibitem{Ball:2022qtp}
R.~D. Ball, A.~Candido, S.~Forte, F.~Hekhorn, E.~R. Nocera, J.~Rojo, and
  C.~Schwan, ``{Parton distributions and new physics searches: the
  Drell\textendash{}Yan forward\textendash{}backward asymmetry as a case
  study},'' \href{http://dx.doi.org/10.1140/epjc/s10052-022-11133-y}{{\em Eur.
  Phys. J. C} {\bfseries 82} no.~12, (2022) 1160},
  \href{http://arxiv.org/abs/2209.08115}{{\ttfamily arXiv:2209.08115
  [hep-ph]}}.

\bibitem{CDF:1995odg}
{\bfseries CDF} Collaboration, F.~Abe {\em et~al.}, ``{Properties of high mass
  multi - jet events at the Fermilab $p\bar{p}$ collider},''
  \href{http://dx.doi.org/10.1103/PhysRevLett.75.608}{{\em Phys. Rev. Lett.}
  {\bfseries 75} (1995) 608--612}.

\bibitem{Huston:1995tw}
J.~Huston, E.~Kovacs, S.~Kuhlmann, H.~L. Lai, J.~F. Owens, D.~E. Soper, and
  W.~K. Tung, ``{Large transverse momentum jet production and the gluon
  distribution inside the proton},''
  \href{http://dx.doi.org/10.1103/PhysRevLett.77.444}{{\em Phys. Rev. Lett.}
  {\bfseries 77} (1996) 444--447},
  \href{http://arxiv.org/abs/hep-ph/9511386}{{\ttfamily arXiv:hep-ph/9511386}}.

\bibitem{Bailey:2019yze}
S.~Bailey and L.~Harland-Lang, ``{Differential Top Quark Pair Production at the
  LHC: Challenges for PDF Fits},''
  \href{http://dx.doi.org/10.1140/epjc/s10052-020-7633-3}{{\em Eur. Phys. J. C}
  {\bfseries 80} no.~1, (2020) 60},
  \href{http://arxiv.org/abs/1909.10541}{{\ttfamily arXiv:1909.10541
  [hep-ph]}}.

\bibitem{Kassabov:2022pps}
Z.~Kassabov, E.~R. Nocera, and M.~Wilson, ``{Regularising experimental
  correlations in LHC data: theory and application to a global analysis of
  parton distributions},''
  \href{http://dx.doi.org/10.1140/epjc/s10052-022-10932-7}{{\em Eur. Phys. J.
  C} {\bfseries 82} no.~10, (2022) 956},
  \href{http://arxiv.org/abs/2207.00690}{{\ttfamily arXiv:2207.00690
  [hep-ph]}}.

\bibitem{DelDebbio:2021whr}
L.~Del~Debbio, T.~Giani, and M.~Wilson, ``{Bayesian approach to inverse
  problems: an application to NNPDF closure testing},''
  \href{http://dx.doi.org/10.1140/epjc/s10052-022-10297-x}{{\em Eur. Phys. J.
  C} {\bfseries 82} no.~4, (2022) 330},
  \href{http://arxiv.org/abs/2111.05787}{{\ttfamily arXiv:2111.05787
  [hep-ph]}}.

\bibitem{Cruz-Martinez:2021rgy}
J.~Cruz-Martinez, S.~Forte, and E.~R. Nocera, ``{Future tests of parton
  distributions},'' \href{http://dx.doi.org/10.5506/APhysPolB.52.243}{{\em Acta
  Phys. Polon. B} {\bfseries 52} (2021) 243},
  \href{http://arxiv.org/abs/2103.08606}{{\ttfamily arXiv:2103.08606
  [hep-ph]}}.

\bibitem{Caola:2022ayt}
F.~Caola, W.~Chen, C.~Duhr, X.~Liu, B.~Mistlberger, F.~Petriello, G.~Vita, and
  S.~Weinzierl, ``{The Path forward to N$^3$LO},'' in {\em {Snowmass 2021}}.
\newblock 3, 2022.
\newblock \href{http://arxiv.org/abs/2203.06730}{{\ttfamily arXiv:2203.06730
  [hep-ph]}}.

\bibitem{Moch:2023tdj}
S.~Moch, B.~Ruijl, T.~Ueda, J.~Vermaseren, and A.~Vogt, ``{Additional moments
  and x-space approximations of four-loop splitting functions in QCD},''
  \href{http://dx.doi.org/10.1016/j.physletb.2024.138468}{{\em Phys. Lett. B}
  {\bfseries 849} (2024) 138468},
  \href{http://arxiv.org/abs/2310.05744}{{\ttfamily arXiv:2310.05744
  [hep-ph]}}.

\bibitem{Cridge:2021pxm}
T.~Cridge, L.~A. Harland-Lang, A.~D. Martin, and R.~S. Thorne, ``{QED parton
  distribution functions in the MSHT20 fit},''
  \href{http://dx.doi.org/10.1140/epjc/s10052-022-10028-2}{{\em Eur. Phys. J.
  C} {\bfseries 82} no.~1, (2022) 90},
  \href{http://arxiv.org/abs/2111.05357}{{\ttfamily arXiv:2111.05357
  [hep-ph]}}.

\bibitem{AbdulKhalek:2019mps}
R.~Abdul~Khalek, S.~Bailey, J.~Gao, L.~Harland-Lang, and J.~Rojo, ``{Probing
  Proton Structure at the Large Hadron electron Collider},''
  \href{http://dx.doi.org/10.21468/SciPostPhys.7.4.051}{{\em SciPost Phys.}
  {\bfseries 7} no.~4, (2019) 051},
  \href{http://arxiv.org/abs/1906.10127}{{\ttfamily arXiv:1906.10127
  [hep-ph]}}.

\bibitem{AbdulKhalek:2019mzd}
{\bfseries NNPDF} Collaboration, R.~Abdul~Khalek, J.~J. Ethier, and J.~Rojo,
  ``{Nuclear parton distributions from lepton-nucleus scattering and the impact
  of an electron-ion collider},''
  \href{http://dx.doi.org/10.1140/epjc/s10052-019-6983-1}{{\em Eur. Phys. J. C}
  {\bfseries 79} no.~6, (2019) 471},
  \href{http://arxiv.org/abs/1904.00018}{{\ttfamily arXiv:1904.00018
  [hep-ph]}}.

\bibitem{NNPDF:2024dpb}
{\bfseries NNPDF} Collaboration, R.~D. Ball {\em et~al.}, ``{Determination of
  the theory uncertainties from missing higher orders on NNLO parton
  distributions with percent accuracy},''
  \href{http://arxiv.org/abs/2401.10319}{{\ttfamily arXiv:2401.10319
  [hep-ph]}}.

\bibitem{Ball:2020xqw}
R.~D. Ball, E.~R. Nocera, and R.~L. Pearson, ``{Deuteron Uncertainties in the
  Determination of Proton PDFs},''
  \href{http://dx.doi.org/10.1140/epjc/s10052-020-08826-7}{{\em Eur. Phys. J.
  C} {\bfseries 81} no.~1, (2021) 37},
  \href{http://arxiv.org/abs/2011.00009}{{\ttfamily arXiv:2011.00009
  [hep-ph]}}.

\bibitem{Ball:2018twp}
{\bfseries NNPDF} Collaboration, R.~D. Ball, E.~R. Nocera, and R.~L. Pearson,
  ``{Nuclear Uncertainties in the Determination of Proton PDFs},''
  \href{http://dx.doi.org/10.1140/epjc/s10052-019-6793-5}{{\em Eur. Phys. J. C}
  {\bfseries 79} no.~3, (2019) 282},
  \href{http://arxiv.org/abs/1812.09074}{{\ttfamily arXiv:1812.09074
  [hep-ph]}}.

\bibitem{Harland-Lang:2018bxd}
L.~A. Harland-Lang and R.~S. Thorne, ``{On the Consistent Use of Scale
  Variations in PDF Fits and Predictions},''
  \href{http://dx.doi.org/10.1140/epjc/s10052-019-6731-6}{{\em Eur. Phys. J. C}
  {\bfseries 79} no.~3, (2019) 225},
  \href{http://arxiv.org/abs/1811.08434}{{\ttfamily arXiv:1811.08434
  [hep-ph]}}.

\bibitem{Ball:2021icz}
R.~D. Ball and R.~L. Pearson, ``{Correlation of theoretical uncertainties in
  PDF fits and theoretical uncertainties in predictions},''
  \href{http://dx.doi.org/10.1140/epjc/s10052-021-09602-x}{{\em Eur. Phys. J.
  C} {\bfseries 81} no.~9, (2021) 830},
  \href{http://arxiv.org/abs/2105.05114}{{\ttfamily arXiv:2105.05114
  [hep-ph]}}.

\bibitem{Kassabov:2022orn}
Z.~Kassabov, M.~Ubiali, and C.~Voisey, ``{Parton distributions with scale
  uncertainties: a Monte Carlo sampling approach},''
  \href{http://dx.doi.org/10.1007/JHEP03(2023)148}{{\em JHEP} {\bfseries 03}
  (2023) 148}, \href{http://arxiv.org/abs/2207.07616}{{\ttfamily
  arXiv:2207.07616 [hep-ph]}}.

\bibitem{AbdulKhalek:2018rok}
R.~Abdul~Khalek, S.~Bailey, J.~Gao, L.~Harland-Lang, and J.~Rojo, ``{Towards
  Ultimate Parton Distributions at the High-Luminosity LHC},''
  \href{http://dx.doi.org/10.1140/epjc/s10052-018-6448-y}{{\em Eur. Phys. J. C}
  {\bfseries 78} no.~11, (2018) 962},
  \href{http://arxiv.org/abs/1810.03639}{{\ttfamily arXiv:1810.03639
  [hep-ph]}}.

\bibitem{Brivio:2017vri}
I.~Brivio and M.~Trott, ``{The Standard Model as an Effective Field Theory},''
  \href{http://dx.doi.org/10.1016/j.physrep.2018.11.002}{{\em Phys. Rept.}
  {\bfseries 793} (2019) 1--98},
  \href{http://arxiv.org/abs/1706.08945}{{\ttfamily arXiv:1706.08945
  [hep-ph]}}.

\bibitem{ZEUS:2019cou}
{\bfseries ZEUS} Collaboration, H.~Abramowicz {\em et~al.}, ``{Limits on
  contact interactions and leptoquarks at HERA},''
  \href{http://dx.doi.org/10.1103/PhysRevD.99.092006}{{\em Phys. Rev. D}
  {\bfseries 99} no.~9, (2019) 092006},
  \href{http://arxiv.org/abs/1902.03048}{{\ttfamily arXiv:1902.03048
  [hep-ex]}}.

\bibitem{Carrazza:2019sec}
S.~Carrazza, C.~Degrande, S.~Iranipour, J.~Rojo, and M.~Ubiali, ``{Can New
  Physics hide inside the proton?},''
  \href{http://dx.doi.org/10.1103/PhysRevLett.123.132001}{{\em Phys. Rev.
  Lett.} {\bfseries 123} no.~13, (2019) 132001},
  \href{http://arxiv.org/abs/1905.05215}{{\ttfamily arXiv:1905.05215
  [hep-ph]}}.

\bibitem{CMS:2021yzl}
{\bfseries CMS} Collaboration, A.~Tumasyan {\em et~al.}, ``{Measurement and QCD
  analysis of double-differential inclusive jet cross sections in proton-proton
  collisions at $ \sqrt{s} $ = 13 TeV},''
  \href{http://dx.doi.org/10.1007/JHEP02(2022)142}{{\em JHEP} {\bfseries 02}
  (2022) 142}, \href{http://arxiv.org/abs/2111.10431}{{\ttfamily
  arXiv:2111.10431 [hep-ex]}}. [Addendum: JHEP 12, 035 (2022)].

\bibitem{Greljo:2021kvv}
A.~Greljo, S.~Iranipour, Z.~Kassabov, M.~Madigan, J.~Moore, J.~Rojo, M.~Ubiali,
  and C.~Voisey, ``{Parton distributions in the SMEFT from high-energy
  Drell-Yan tails},'' \href{http://dx.doi.org/10.1007/JHEP07(2021)122}{{\em
  JHEP} {\bfseries 07} (2021) 122},
  \href{http://arxiv.org/abs/2104.02723}{{\ttfamily arXiv:2104.02723
  [hep-ph]}}.

\bibitem{McCullough:2022hzr}
M.~McCullough, J.~Moore, and M.~Ubiali, ``{The dark side of the proton},''
  \href{http://dx.doi.org/10.1007/JHEP08(2022)019}{{\em JHEP} {\bfseries 08}
  (2022) 019}, \href{http://arxiv.org/abs/2203.12628}{{\ttfamily
  arXiv:2203.12628 [hep-ph]}}.

\bibitem{Gao:2022srd}
J.~Gao, M.~Gao, T.~J. Hobbs, D.~Liu, and X.~Shen, ``{Simultaneous CTEQ-TEA
  extraction of PDFs and SMEFT parameters from jet and $ t\overline{t} $
  data},'' \href{http://dx.doi.org/10.1007/JHEP05(2023)003}{{\em JHEP}
  {\bfseries 05} (2023) 003}, \href{http://arxiv.org/abs/2211.01094}{{\ttfamily
  arXiv:2211.01094 [hep-ph]}}.

\bibitem{Kassabov:2023hbm}
Z.~Kassabov, M.~Madigan, L.~Mantani, J.~Moore, M.~Morales~Alvarado, J.~Rojo,
  and M.~Ubiali, ``{The top quark legacy of the LHC Run II for PDF and SMEFT
  analyses},'' \href{http://dx.doi.org/10.1007/JHEP05(2023)205}{{\em JHEP}
  {\bfseries 05} (2023) 205}, \href{http://arxiv.org/abs/2303.06159}{{\ttfamily
  arXiv:2303.06159 [hep-ph]}}.

\bibitem{Hammou:2023heg}
E.~Hammou, Z.~Kassabov, M.~Madigan, M.~L. Mangano, L.~Mantani, J.~Moore, M.~M.
  Alvarado, and M.~Ubiali, ``{Hide and seek: how PDFs can conceal new
  physics},'' \href{http://dx.doi.org/10.1007/JHEP11(2023)090}{{\em JHEP}
  {\bfseries 11} (2023) 090}, \href{http://arxiv.org/abs/2307.10370}{{\ttfamily
  arXiv:2307.10370 [hep-ph]}}.

\bibitem{Costantini:2024xae}
M.~N. Costantini, E.~Hammou, Z.~Kassabov, M.~Madigan, L.~Mantani,
  M.~Morales~Alvarado, J.~M. Moore, and M.~Ubiali, ``{SIMUnet: an open-source
  tool for simultaneous global fits of EFT Wilson coefficients and PDFs},''
  \href{http://arxiv.org/abs/2402.03308}{{\ttfamily arXiv:2402.03308
  [hep-ph]}}.

\bibitem{Shen:2024uop}
X.-M. Shen, J.~Gao, M.~Gao, T.~J. Hobbs, and D.~Liu, ``{Determining SMEFT and
  PDF parameters simultaneously based on the CTEQ-TEA framework},''
  \href{http://dx.doi.org/10.22323/1.449.0291}{{\em PoS} {\bfseries
  EPS-HEP2023} (2024) 291}.

\bibitem{Forte:2020pyp}
S.~Forte and Z.~Kassabov, ``{Why $\alpha _s$ cannot be determined from hadronic
  processes without simultaneously determining the parton distributions},''
  \href{http://dx.doi.org/10.1140/epjc/s10052-020-7748-6}{{\em Eur. Phys. J. C}
  {\bfseries 80} no.~3, (2020) 182},
  \href{http://arxiv.org/abs/2001.04986}{{\ttfamily arXiv:2001.04986
  [hep-ph]}}.

\bibitem{Iranipour:2022iak}
S.~Iranipour and M.~Ubiali, ``{A new generation of simultaneous fits to LHC
  data using deep learning},''
  \href{http://dx.doi.org/10.1007/JHEP05(2022)032}{{\em JHEP} {\bfseries 05}
  (2022) 032}, \href{http://arxiv.org/abs/2201.07240}{{\ttfamily
  arXiv:2201.07240 [hep-ph]}}.

\bibitem{Torre:2020aiz}
R.~Torre, L.~Ricci, and A.~Wulzer, ``{On the W\&Y interpretation of high-energy
  Drell-Yan measurements},''
  \href{http://dx.doi.org/10.1007/JHEP02(2021)144}{{\em JHEP} {\bfseries 02}
  (2021) 144}, \href{http://arxiv.org/abs/2008.12978}{{\ttfamily
  arXiv:2008.12978 [hep-ph]}}.

\bibitem{NNPDF:2021uiq}
{\bfseries NNPDF} Collaboration, R.~D. Ball {\em et~al.}, ``{An open-source
  machine learning framework for global analyses of parton distributions},''
  \href{http://dx.doi.org/10.1140/epjc/s10052-021-09747-9}{{\em Eur. Phys. J.
  C} {\bfseries 81} no.~10, (2021) 958},
  \href{http://arxiv.org/abs/2109.02671}{{\ttfamily arXiv:2109.02671
  [hep-ph]}}.

\bibitem{AbdulKhalek:2021gbh}
R.~Abdul~Khalek {\em et~al.}, ``{Science Requirements and Detector Concepts for
  the Electron-Ion Collider: EIC Yellow Report},''
  \href{http://arxiv.org/abs/2103.05419}{{\ttfamily arXiv:2103.05419
  [physics.ins-det]}}.

\bibitem{LHeC:2020van}
{\bfseries LHeC, FCC-he Study Group} Collaboration, P.~Agostini {\em et~al.},
  ``{The Large Hadron-Electron Collider at the HL-LHC},''
  \href{http://dx.doi.org/10.1088/1361-6471/abf3ba}{{\em J. Phys. G} {\bfseries
  48} no.~11, (2021) 110501}, \href{http://arxiv.org/abs/2007.14491}{{\ttfamily
  arXiv:2007.14491 [hep-ex]}}.

\bibitem{Cruz-Martinez:2023sdv}
J.~M. Cruz-Martinez, M.~Fieg, T.~Giani, P.~Krack, T.~M\"akel\"a,
  T.~Rabemananjara, and J.~Rojo, ``{The LHC as a Neutrino-Ion Collider},''
  \href{http://arxiv.org/abs/2309.09581}{{\ttfamily arXiv:2309.09581
  [hep-ph]}}.

\end{thebibliography}

\providecommand{\href}[2]{#2}\begingroup\raggedright\endgroup

\end{document}